\documentclass[11pt]{article}

\usepackage{amsmath,amssymb,amsthm,graphicx,latexsym}
\usepackage{hyperref}
\usepackage{graphics}
\usepackage{setspace}
\usepackage{xcolor}
\newcommand{\ed}{\end{document}}
\newcommand{\beq}{\begin{equation}}
\newcommand{\eeq}{\end{equation}}
\newcommand{\beqa}{\begin{eqnarray}}
\newcommand{\eeqa}{\end{eqnarray}}
\newcommand{\bc}{\begin{center}}
\newcommand{\ec}{\end{center}}

\newcommand{\ba}{\begin{array}}
\newcommand{\ea}{\end{array}}

\begin{document}

\title{{\bf{Relativistic Fluid Dynamics in Curved Spacetime:\\ a Novel Effective Hamiltonian Approach
 }}}
	\author{  {\bf {\normalsize Arpan Krishna Mitra$^{1, 3}$}$
			$\thanks{E-mail: arpankmitra@gmail.com}}$~$
		{\bf {\normalsize Subir Ghosh$^2$}$
			$\thanks{E-mail: subir\_ ghosh2@rediffmail.com}},$~$
		\\{\normalsize $^1$Aryabhatta Research Institute of Observational Sciences}
		\\{\normalsize Manora peak, Beluwakhan, Uttarakhand 263002}
		\\\\
		{\normalsize $^2$Indian Statistical Institute}
		\\{\normalsize  203, Barrackpore Trunk Road, Kolkata 700108, India}
        \\\\
        \\{\normalsize $^3$Harish Chandra Research Institute, HBNI,}
		\\{\normalsize Chhatnag Rd, Jhusi, Uttar Pradesh 211019}
		\\[0.3cm]
}
	\date{}
	\maketitle
\begin{abstract}
 
We present a comprehensive Eulerian (Hamiltonian) framework for the analysis of relativistic fluid dynamics in arbitrarily curved spacetimes, with particular emphasis on Schwarzschild geometry. The principal innovation of this work lies in the systematic use of density and three-dimensional velocity fields, all defined in coordinate time (or Newtonian time), while rigorously incorporating relativistic and spacetime curvature effects throughout the formulation. {\it{We stress that the dynamics is developed in coordinate (Eulerian) time, as seen from the perspective of an Eulerian observer, placed at a fixed position}}.\\
In the non-relativistic regime, it is well-established that Poisson brackets between the density $\rho(\vec{x}, t)$ and velocity fields $v^i(\vec{x}, t)$, combined with an appropriate Hamiltonian, yield the continuity and Euler equations as Hamiltonian equations of motion. These field-theoretic Poisson brackets can be derived from canonical discrete phase space Poisson brackets, originally defined on Lagrangian variables $(x_i, p_j)$, via a mapping to Eulerian field variables $(\rho(\vec{x}, t), v^i(\vec{x}, t))$.\\
Building on this foundation, we develop a natural generalisation of the discrete system, consisting of the Poisson brackets and Hamiltonian, to its relativistic counterpart in curved spacetime. By leveraging the same variable mapping, we construct a relativistic version of the Eulerian Poisson algebra for $(\rho, v^i)$ fields and propose a corresponding fluid Hamiltonian. This framework enables the derivation of relativistic continuity and Euler equations that are applicable to generic curved spacetimes, with all dynamical evolution formulated explicitly in coordinate time.\\
Focusing on the Schwarzschild metric, we obtain explicit, stationary solutions for purely radial background flows and perform a detailed stability analysis via perturbative techniques. Furthermore, we explore how a generalised notion of vorticity influences perturbations superimposed on an irrotational (zero vorticity) background, revealing the subtle role of vorticity in the stability of relativistic fluid flows in curved geometries.\\
This work provides a unified, first-principles Hamiltonian approach to relativistic fluid dynamics in curved spacetimes and establishes a foundation for further investigations into astrophysical and cosmological fluid phenomena.
\end{abstract}

\section{ Introduction to Eulerian fluid dynamics:}  Fluid dynamics, one of the oldest branches of physical science \cite{LandauLifshitz, Weinberg, Eckart, BrownActionFluid,  lamb,  Galilei2023, An2024, MorrisonRMP, Morrison82, Arnold}, makes its presence strongly felt in diverse areas of applied and theoretical fields of science. Its relevance in modern-day high-energy physics rests on the fact that it can be described in general terms as a universal
description of long-wavelength/low-energy physics. It can be considered the low-energy effective model of an exact theory where the degrees of freedom of the fluid model are density and velocity fields, obtained from a mapping between the degrees of freedom of the exact theory and the fluid field variables. The physical reasoning behind the success of fluid models is the assumption that at sufficiently high energy densities, local
equilibrium prevails in an interacting field theory so that local inhomogeneities are smoothed out enough
to reduce to continuous fluid variables.   In the fluid picture, the dynamics is governed by the
continuity equation and the Euler force equation \cite{ lamb}. We will focus on the Hamiltonian framework where these equations are derived as Hamilton's equations of motion, using a Hamiltonian and Poisson Brackets (PB) \cite{MorrisonRMP, Morrison82, Arnold, Schutz, AnderssonComer, pb, CanonicalNote2024, mitra2018, Sanders2024, Holm19851, dix45}. (The action principle in the present case bears some subtlety due to the singular nature of the symplectic structure (for a discussion see \cite{jac}) that requires a first-order formalism with Clebsch variables \cite{lim, HolmKupershmidt1983}. We will closely follow the framework, elaborated extensively in \cite{jac}, where the fluid (Euler) variable brackets are derived from discrete  (Lagrangian) canonical coordinate-momentum Poisson brackets, using a map that connects the Euler and Lagrange degrees of freedom. The existing formalism is purely non-relativistic and lives in a flat 3-dimensional space.

In the present work, we will study the fascinating topic of fluid dynamics in extreme spacetimes, such as Black Holes (BH) \cite{Michel1972, MTW, Wald, Padmanabhan, Chandrasekhar, Pretorius, bondi, DasAASarkar2001, Das_GRWind2002, Das_PseudoSchwarzschildOutflow2002, MandalRayDas_MNRAS2007}. Obviously, this will require a non-trivial extension of the above-mentioned scheme, which we have developed here from scratch. This will require a generalisation of the Hamiltonian and PB structure that are consistent with a relativistic curved spacetime \cite{ NoncanonicalReview2025, NSF2024, daSilva2024, OlsthoornReview, AnderssonComer2021, Fajman2024, PhysRevD.91.064055, Holm1985, LichnerowiczHydro}. The novelty and advantage of our approach is that the basic variables remain the same as in the non-relativistic case, {\it{i.e.}} density and velocity fields and the dynamics is enacted in conventional (Newtonian) time. The relativistic and spacetime curvature effects appear as corrections in the continuity and  Euler equations. The present scheme will be applicable to any stationary metric, but in the present paper, we will further restrict the application to static metric, in particular, the Schwarzschild metric. The most important point and novelty of our scheme is that we will focus on the coordinate velocity, also technically known as Eulerian velocity, and furthermore, the analysis will be from the perspective of an Eulerian observer, fixed at a particular position. 

The conventional way is to postulate the covariant form of energy-momentum tensor for an ideal fluid  $T_{\mu\nu}=(p+\rho)U_\mu U_\nu -p\eta_{\mu\nu}$ \cite{LandauLifshitz} with the Einstein equation and conservation equation given by
\begin{equation}
R_{\mu\nu}-g_{\mu\nu}R=T_{\mu\nu},~~\nabla _\mu T^{\mu\nu}=0.
\label{1}
\end{equation}
where $R_{\mu\nu}, R$ are the Ricci tensor and scalar and  $p,\rho, U_\mu$ are pressure, density and four-velocity of the fluid, respectively. The dynamical evolution of fluid proceeds in proper time or along flow lines. However,  this setup obscures direct physical interpretations for static Eulerian observers and is not suitable for a  Hamiltonian formulation.   \\
 Contrary to the above (instead of working in a manifestly covariant setup), we will consider a space $\bigoplus$ time broken-up scenario. Our approach retains the \textit{Eulerian description}---using coordinate time as the evolution parameter and treating density and spatial velocity as fundamental variables,  in a relativistic curved spacetime setting. This enables us to:
\begin{itemize}
    \item Maintain a Hamiltonian structure with explicit Poisson brackets,
    \item Keep the formalism closer to non-relativistic intuition, where density and velocity fields evolve under Hamiltonian flow.
     \item Facilitate numerical simulations and perturbative analysis \cite{VillaRampf2015, Friedman2013,DiazGuerra2024}, especially near horizons, in a frame fixed to a particular coordinate system (e.g., Schwarzschild coordinate time).
    \item Provide a real-time description of the fluid behaviour as seen by a distant static observer ({\it{e.g.}}  on earth).
\end{itemize}
We stress that this alternative is not just one more computational prescription: it shifts the observational perspective to that of a coordinate (Eulerian) observer, who measures physical quantities at fixed spatial points in the global spacetime chart---an approach that aligns with most astrophysical measurements made from asymptotic infinity \cite{bondi, DasAASarkar2001, Das_GRWind2002, Das_PseudoSchwarzschildOutflow2002, MandalRayDas_MNRAS2007, Johnston_deVegt, Kopeikin_ref, Gaia_EDR3}.

\section{Fluid dynamics in Hamiltonian framework: kinematics} {\it{Non-relativistic scenario}}:  Let us briefly recapitulate the non-relativistic Eulerian fluid dynamics in flat spacetime. One starts by postulating a (intuitively expected form of)  Hamiltonian \cite{jac}
\begin{equation}
H=\frac{1}{2}\int d^3x~\rho(x)\vec {v}^2(x)
    \label{2a}
\end{equation}
with a PB structure between density $\rho(\vec x)$ and velocity $v^i(\vec x)$ fields
\begin{equation}
\{\rho(\vec x),\rho(\vec y)\}=0,~~\{\rho(\vec x),v^i(\vec y)\}=\partial^i_x\delta(\vec x-\vec y),~~\{v^i(\vec x),v^i(\vec y)\}=\frac{\omega^{ij}}{\rho}\delta(\vec x-\vec y)
    \label{pp1}
\end{equation}
where  $\omega^{ij}=\partial^iv^j-\partial^jv^i$ is known as the vorticity \cite{jac}. The continuity and Euler equations are obtained from 
\begin{equation}
\dot \rho (\vec r)=\{\rho(\vec r),H\}=-\partial_i(\rho v^i), ~~\dot v^i (\vec r)=\{v^i(\vec r),H\}=-v_j\partial^jv^i.
    \label{3}
\end{equation}
Crucial to our work is that the above PB (\ref{pp1}) can be computed from the basic canonical PB
\begin{equation}
\{X^i(\vec x),X^j(\vec y)\}=\{P_i(\vec x),P_j(\vec y)\}=0,$$$$\{X^i(\vec x),P_j(\vec y)\}=\{X^i(\vec x),\rho_0 U_j(\vec y)\}=\eta^i_j\delta(\vec x-\vec y).
    \label{d4}
\end{equation}
with $\rho, v^i, j^i=\rho v^i$ defined by the constitutive relations
\begin{equation}
\rho (\vec r,t) =\rho_0\int d^3x~\delta (\vec X(\vec x)-\vec r),~j^i=\rho(\vec r)v^i(\vec r,t)= \rho_0\int d^3x~\dot {X}^i(\vec x)\delta (\vec X(\vec x)-\vec r) 
    \label{d5}
\end{equation}
with $\rho_0$ is a dimensional constant denoting a background density \cite{jac}. 
The $\vec X(\vec x),\vec U(\vec x)$ represent the microscopic (Lagrangian) degrees of freedom of the fluid particles, extended to a many body (strictly speaking an innumerable) continuum. We have dropped the time argument since PBs and all PB analysis are at equal time.\\
{\it{Relativistic field algebra}}: The covariant form of canonical PB is given by \cite{dix45}
 \begin{equation}
\{X^\mu, X^\nu \}=0, ~\{X^\mu, P_\nu \}=g^\mu_\nu =\eta^\mu_\nu,~\{P_\mu, P_\nu \}=0 .
     \label{d1} 
 \end{equation}
  This is consistent with a covariant particle action given by
 \begin{equation}
S=\int \frac{m}{2}g_{\mu\nu}(X^\lambda )\frac{dX^\mu}{d\tau}\frac{dX^\nu}{d\tau}
    \label{xd1}
\end{equation}
with  canonical momenta \cite{bert}
 \begin{equation}
P_\mu=\frac{\partial L}{\ \partial (dX^\mu/d\tau )}=mg_{\mu\nu}\frac{dX^\nu}{d\tau}.
     \label{2}
 \end{equation}
 Note that in order to exploit the mapping (\ref{d5})  and subsequent field algebra (\ref{pp1}) in relativistic case, we need to generalize the  PB for $X^i,{\dot X}^j$ at time $t$ to their covariant and curved spacetime counterparts $X^\mu, U^\nu\equiv dX^\nu/dt$ . The above brackets in (\ref{d1}) reproduce 
  \begin{equation}
g^{\alpha\nu}\{X^\mu, P_\nu \}=g^{\alpha\nu}g^\mu_\nu=g^{\alpha\mu}\rightarrow \{X^\mu, \frac{dX^\nu}{d\tau} \}=\frac{1}{m}g^{\mu\nu},
     \label{4}
 \end{equation}
  \begin{equation}
\{\frac{dX^\alpha}{d\tau},\frac{dX^\beta}{d\tau}\}=\frac{1}{m^2}\{g^{\alpha\mu}P_\mu,g^{\beta\nu}P_\nu \}=
  \frac{1}{m}(g^{\beta\mu}\partial_\mu g^{\alpha\nu}-g^{\alpha\mu}\partial_\mu g^{\beta\nu})g_{\nu\sigma}\frac{dX^\sigma}{d\tau}
     \label{5}
 \end{equation}
 where the effect of curved metric $g^{\mu\nu}$ is manifest. 
 Note the $X^\mu, dX^\nu/d\tau$ satisfy all the non-trivial Jacobi identities \cite{Arnold}.
Following \cite{jac}, we will use only contravariant components. Again, this is a nontrivial requirement because of the curved metric. 

Next task is to substitute time $t$  in place of the proper time $\tau$. From the definition 
  \begin{equation}
d\tau^2=-g_{\mu\nu}dX^\mu dX^\nu 
     \label{6}
 \end{equation}
we recover
  \begin{equation}
ds=dt {\sqrt{-(g_{00}+2g_{0i}U^i+g_{ij}U^iU^j)}} ~~\equiv \Phi dt   
     \label{7}
 \end{equation} 
 where we define velocity $U^\mu=\frac{dX^\mu}{dt}$ and $$ ~dt=dX^0,~U^0=1, ~U^i=dX^i/dt .$$
 Finally we obtain the cherished form of algebra
\begin{equation}
\{X^\mu,U^\nu\}=\frac{\Phi}{m} g^{\mu\nu}
     \label{10}
 \end{equation}
 \begin{equation}
\{U^\alpha,U^\beta\}=\frac{\Phi}{m}(g^{\beta\mu}\partial_\mu g^{\alpha\nu}-g^{\alpha\mu}\partial_\mu g^{\beta\nu})g_{\nu\sigma}U^\sigma .
     \label{11}
 \end{equation}
From now on, we will introduce a low velocity approximation such that $\Phi \approx {\sqrt{-g_{00}}}$, primarily to contain the non-linearity involving $U^i$. This set of PB can be generalised to field theory similar to the non-relativistic case discussed earlier (see (\ref{d4})). 

Using identical mapping as in (\ref{d5}) for the the non-relativistic setup ,  the  density $\rho =J^0$ and current fields $J^i$,  are defined as
$$
J^0(\vec r,t)=\rho (\vec r,t)=\rho_0\int d^3x ~\delta (\vec X(\vec x,t)-\vec r ),~~
j^i(\vec r,t)=\rho_0\int d^3x~ U^i(\vec x,t)\delta (\vec X(\vec x,t)-\vec r ) .
   $$
to compute the generalised $\rho, J^i$ current algebra with relativistic and curved spacetime effects,
\begin{equation}
\{\rho(\vec r),\rho(\vec{r}'\}=0,~~ \{j^i(\vec r),\rho(\vec{r}'\}=-\Phi(\vec r)\rho(\vec r)g^{ij}(\vec r)\partial_j\delta (\vec r-\vec r'), 
    \label{j3}
\end{equation}
\begin{equation}
\{j^i(\vec r),j^j(\vec{r}'\}=-\Phi(\vec r)g^{il}(\vec r)j^j(\vec r)\partial_l\delta (\vec r-\vec r')+\Phi(\vec r')g^{jl}(\vec r')j^i(\vec r'){\partial}'_l\delta (\vec r-\vec r') $$$$
+\Phi (g^{j\mu}\partial_\mu g^{i\nu}-g^{i\mu}\partial_\mu g^{j\nu})g_{\nu\sigma}j^\sigma \delta (\vec r-\vec r')
    \label{j4}
\end{equation}
with $$\Phi={\sqrt{-(g_{00}+2g_{0i}u^i+g_{ij}u^iu^j)}}\approx {\sqrt{-g_{00}}}.$$
Exploiting the identification $j^i(\vec r)=\rho(\vec r)U^i(\vec r)$ an equivalent brackets structure is given by
\begin{equation}
\{U^i(\vec r),\rho(\vec{r}'\}=-\Phi(\vec r)g^{ij}(\vec r)\partial_j\delta(\vec r -\vec{r}'),~~\{U^i(\vec r),U^j(\vec{r}'\}=\frac{\Phi}{\rho}(g^{ik}\partial_k U^j-g^{jk}\partial_k U^i)\delta(\vec r -\vec{r}')
\label{x}
\end{equation}
where $\Omega^{ij}=(g^{ik}\partial_k U^j-g^{jk}\partial_k U^i)$ is  the generalized vorticity  \cite {jac, NoncanonicalReview2025, daSilva2024, PhysRevD.91.064055, Holm1985, LichnerowiczHydro}of the fluid. As expected, this the PB algebra (\ref{pp1}) is recovered once $c\rightarrow \infty,~g^{\mu\nu}\rightarrow \eta^{\mu\nu}$ limits are imposed. The relativistic and curved spacetime extension of fluid variable algebra in (\ref{j3}, \ref{j4}, \ref{x}) constitutes our first set of major results.   \\
 
{\it{Relativistic fluid Hamiltonian}}: The remaining task is to write down the Hamiltonian in $(X^i,P_j)$-phase space or equivalently in $(X^i,U^j)$ space, which is obtained by identifying $H(P_j,S^j,t)=-P_0=-\frac{m}{\Phi}g_{0\nu}u^\nu$ (recall that $P_\mu=\frac{m}{\Phi}g_{\mu\nu}U^\nu$) and solving for $P_0$ using the dispersion relation $g^{\mu\nu}P_\mu P_\nu=-m^2$ \cite{bert},
  \begin{equation}
H=-P_0=\frac{g^{0i}P_i}{g^{00}} +[\frac{g^{ij}P_iP_j+m^2}{-g^{00}}+(\frac{g^{0i}P_i}{g^{00}})^2]^{1/2} $$$$
=\frac{mg^{0i}U_i}{g^{00}} +[\frac{m^2(g^{ij}U_i U_j+1)}{-g^{00}}+(\frac{mg^{0i}U_i}{g^{00}})^2]^{1/2}.
    \label{a8}
\end{equation}
The corresponding field theoretic Hamiltonian is 
\begin{equation}
H=\int d^3r \frac{\rho}{g^{00}}(g^{0i}g_{i\alpha}U^\alpha +{\sqrt{\Psi}}), $$$$
{\sqrt{\Psi}}=(g^{ij}g_{i\alpha}g_{j\beta}U^\alpha U^\beta +1)+(g^{0i}g_{i\alpha} U^\alpha)^2
  \label{psi}
  \end{equation}
  which reduces to the correct non-relativistic and flat spacetime limit. Note that for a diagonal metric and low $U^i$ ($\sim $ non-relativistic limit) the Hamiltonian is given by
  $$H\approx \int~d^3r (\frac{\rho}{2g^{00}}~g_{ij}U^iU^j +\frac{\rho}{g^{00}})$$
  where the effective potential (in Newtonian terminology) is $V(\rho)=\rho/g^{00}$, corresponding to a (pressure $P=\rho (dV/d\rho) -\rho =0$) pressure-less fluid, equivalent to dust. \footnote{ To introduce a non-zero barotropic pressure, one can modify the $\Psi$-term by $\psi=g^{ij}g_{i\alpha}g_{j\beta}U^\alpha U^\beta +\nu V(\rho)$, $\nu$ being a constant. We have not used this option in the present work.}
  The generalized fluid Hamiltonian and fluid algebra concludes first part of our work. The next section will be devoted to generate the generalized fluid dynamics, in Hamiltonian framework.
\section{Fluid dynamics (in coordinate time) in Hamiltonian framework: \\
background flow}
After a long and tedious computation we recover the fluid equations of motion for relativistic regime and in curved spacetime,
\begin{equation}
\dot \rho =\{H,\rho\}=\partial_n \bigg{[}  \frac{\Phi\rho}{g^{00}}g^{ln}\bigg{(} g^{0i}g_{il}+\frac{1}{\sqrt{\chi}}
(g^{0i}g^{0j}-g^{00}g^{ij})g_{jl}g_{i\alpha}U^\alpha \bigg{)}\bigg{]}
    \label{v1}
\end{equation}

\begin{equation}
\dot{U}^i=\{H,{U}^i\}=\Phi g^{ik}\partial_k\bigg{[} \frac{1}{g^{00}}\bigg{(}\frac{g^{0l}}{g^{00}}(g_{l0}+g_{lj}U^j)+\sqrt{\chi}\bigg{)}\bigg{]} $$$$
-\frac{\Phi}{(g^{00})^2}\omega^{ij}\bigg{[}g^{0k}g_{kj}-\frac{1}{\sqrt{\chi}}\bigg{(}g_{lj}(g^{lk}g_{k0}+g^{nl}g_{nk}U^k)-\frac{g^{0n}g_{nj}}{g^{00}}(g^{0r}g_{0r}+g^{0k}g_{kl}U^l)\bigg{)}\bigg{]}
    \label{v2}
\end{equation}
\begin{equation}
\dot{U}^i=\Phi g^{ik}\partial_k\bigg{(}\frac{1}{g^{00}}(g^{0j}g_{j\alpha}U^\alpha +\sqrt{\chi}~)\bigg{)}$$$$
-\frac{\Phi}{g^{00}}\Omega^{ij}\bigg{(}g^{0k}g_{kj}+\frac{1}{\sqrt{\chi}}(g^{0m}g^{0k}-g^{00}g^{mk})g_{kj}g_{m\alpha}U^\alpha \bigg{)}.
    \label{v4}
\end{equation}
The above are applicable to any stationary metric. 
For a diagonal (static) metric the above equations simplify considerably,
\begin{equation}
\dot\rho +\partial_i\bigg{(}\frac{\Phi\rho}{\sqrt{\chi}}U^i\bigg{)}=0 \Longrightarrow 
    \dot\rho +\partial_i\bigg{(}\frac{(-g_{00})\rho}{\sqrt{1+g_{ij}U^iU^j}}U^i\bigg{)}=0
    \label{dd1}
\end{equation}
where $\chi=-g^{00}(1+g_{ij}U^iU^j),~~\Phi \approx {\sqrt{-g_{00}}}$ and
\begin{equation}
\dot{U}^i=-{\sqrt{-g_{00}}}g^{ij}\partial_j\bigg{(}{\sqrt{-g_{00}(1+g_{kl}U^kU^l)}}\bigg{)}-\frac{g_{00}}{{\sqrt{(1+g_{kl}U^kU^l)}}}\omega^{ij}g_{jk}U^k
    \label{v6}
\end{equation}
For non-relativistic system in  flat spacetime we recover the standard equations
$$\dot \rho +\partial_i(\rho U^i)=0, ~~\dot{U}^i=-U_j\partial^jU^i .$$
Using the abbreviations
$$
Z = g_{rr}(U^r)^2 + g_{\theta\theta}(U^\theta)^2 + g_{\phi\phi}(U^\phi)^2 ,~~
\Phi = \sqrt{ -g_{00}(1 + Z) }$$
the resultant form of the fluid equations for a generic {\it{diagonal}} metric is
\begin{equation}
 \dot\rho +\partial_i\bigg{(}\frac{-g_{00})\rho U^i}{\sqrt{1+Z}}\bigg{)}=0, ~\dot{U}^i = - \sqrt{-g_{00}}\, g^{ij} \partial_j\Phi 
- \frac{g_{00}}{ \sqrt{ 1 + Z } }\, \omega^{ij} g_{jk} U^k .
\end{equation}
This is one of our primary major results. Neglecting higher orders of $U^i$, the equations simplify to 
\begin{equation}
 \dot\rho +\partial_i((-g_{00})\rho U^i )=0, ~\dot{U}^i = - \sqrt{-g_{00}}\, g^{ij} \partial_j\Phi 
- g_{00}\, \omega^{ij} g_{jk} U^k
\end{equation}
which will be our working equations. As we are interested in considering the background to be Schwarzschild spacetime, we will rewrite the equations in spherical polar coordinates $r,\theta,\phi$. \\
{\it{Schwarzschild background}}: Specializing to Schwarzschild spacetime \cite{Chandrasekhar, Pretorius, Benomio2022}, with the metric  parameterized by
\begin{equation}
    g_{\mu\nu} =
    \begin{bmatrix}
    -f(r) & 0 & 0 & 0 \\
    0 & f(r)^{-1} & 0 & 0 \\
    0 & 0 & r^2 & 0 \\
    0 & 0 & 0 & r^2 \sin^2\theta
    \end{bmatrix}, \quad f(r) = 1 - \frac{B}{r}, \quad B = 2GM
\end{equation}
the equations, (up to $O|U|^2$),  simplify to
\begin{align}
\dot{U}^r =-\ & \frac{f}{2} \left[ \frac{B}{r^2}(1 + Z) + \partial_r Z \right] \nonumber \\
& - f \left( \omega^{r\theta} r^2 U^\theta + \omega^{r\phi} r^2 \sin^2\theta\, U^\phi \right),
\label{x11}
\end{align}

\begin{align}
\dot{U}^\theta =-\ & \frac{ {f} }{2\, r^2 }   \, \partial_\theta Z  \nonumber \\
& - f \left( \omega^{\theta r} \frac{U^r}{f} + \omega^{\theta\phi} r^2 \sin^2\theta\, U^\phi \right),
\label{x12}
\end{align}

\begin{align}
\dot{U}^\phi =-\ & \frac{ f}{2\, r^2 \sin^2\theta } \, \partial_\phi Z \nonumber \\
& - f \left( \omega^{\phi r} \frac{U^r}{f} + \omega^{\phi\theta} r^2\, U^\theta \right).
\label{x13}
\end{align}
The continuity equation is  given by 
\begin{equation}
\dot{\rho}
\;+\;
\frac{1}{r^2}\,\frac{\partial}{\partial r}\Biggl[
r^2 f(r)\rho U^r \Biggr]+ \frac{1}{r\,\sin\theta}\,\Biggl[\frac{\partial}{\partial\theta}
\sin\theta
f(r)\rho U^\theta
\Biggr] + \frac{1}{r\,\sin\theta}\,\frac{\partial}{\partial\phi}\Biggl[
f(r)\rho
\,U^\phi
\Biggr]
=0.
\end{equation}


{\it{Spherically symmetric background flow}}: We consider a spherically symmetric fluid system in Schwarzschild spacetime \cite{Michel1972, Wald} so that only the radial component of fluid velocity $U^r$ is non-zero and the variables can depend only on $t,r$. We use the notation $\rho_0(r,t),~U^r\equiv U_0(r,t),~ U^\theta\equiv V(r,t),~ U^\phi\equiv W(r,t)$ and furthermore $V=W=0$. Hence the background profile at $t=0$ is $\rho_0(r,t),~U_0(r,t),V=0,W=0$. This leads to the continuity and Euler  equations, respectively, for the background flow as 
\begin{equation}
\dot{\rho}_0 + \frac{1}{r^2}\,\frac{\partial}{\partial r}\Biggl[
r^2 f(r)\rho U_{0} \Biggr]=0\approx \dot{\rho}_0 + \frac{1}{r^2}\,\frac{\partial}{\partial r}\Biggl[
r^2 \Bigg(1 - \frac{B}{r}\Bigg)\rho U_{0} \Biggr]=0
\end{equation}

\begin{equation}
 \dot{U}_0 =- \frac{f^{3/2}}{2 \Phi} \left[ \frac{B}{r^2}(1 + Z) + \partial_r Z \right]\approx - \frac{f}{2}\bigg{[}( 1+\frac{U_0^2}{f})\frac{B}{r^2}+\partial_r(\frac{U_0^2}{f})  \bigg{]} 
\label{l1}
\end{equation}
where the vorticity terms obviously drop out. The $U^\theta, U^\phi$ equations reduce to trivial $0=0$ identities. We have restricted the equations up to $O(U_0^2)$ terms.

\begin{figure}
    \centering
    \includegraphics[width=0.9\linewidth]{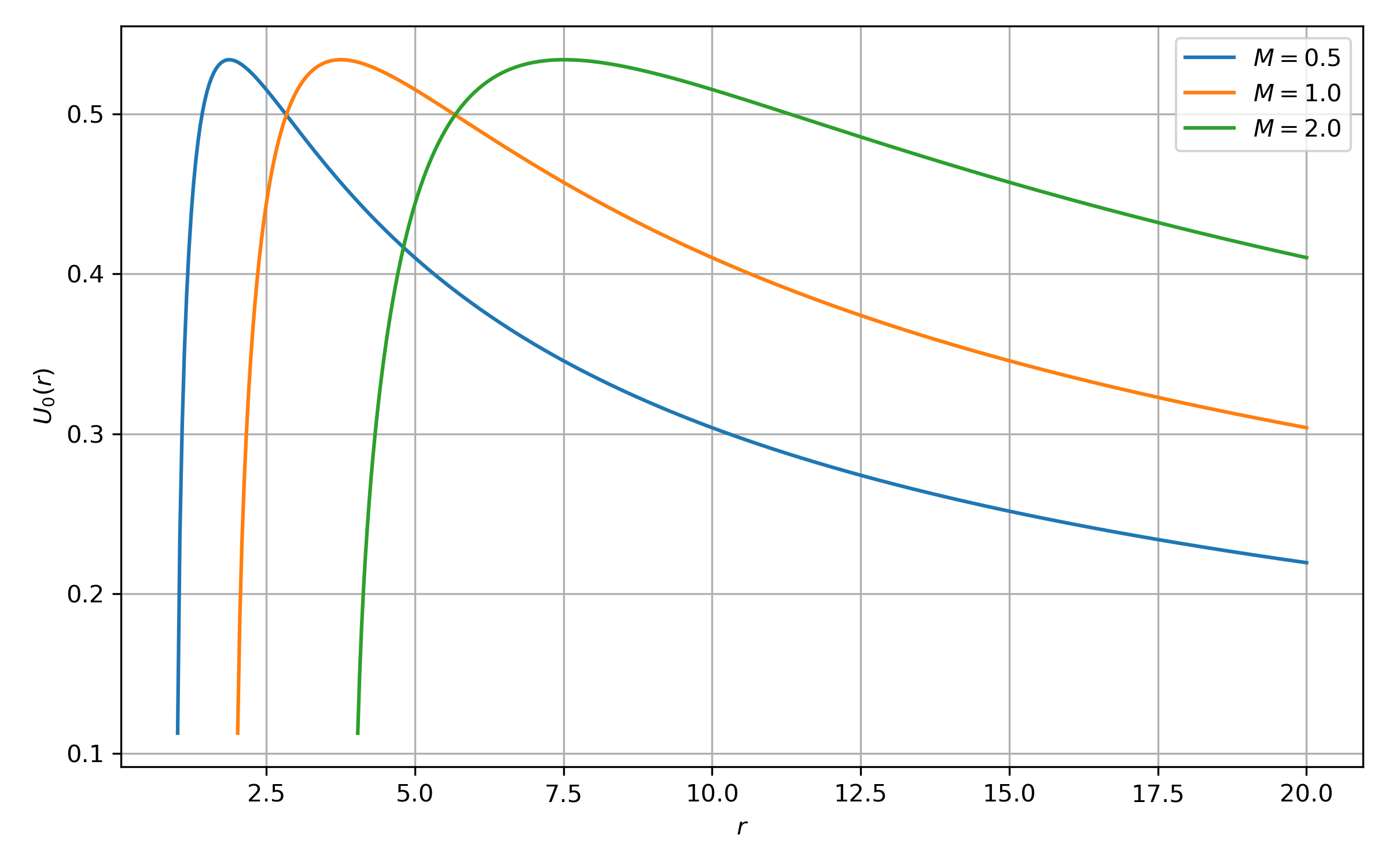}
    \caption {Outward stationary background flow velocity $U_{0}$ vs $r$ for different black hole mass $M$ }
    \label{fig:enter-theoq1}
\end{figure}

\begin{figure}
    \centering
    \includegraphics[width=0.9\linewidth]{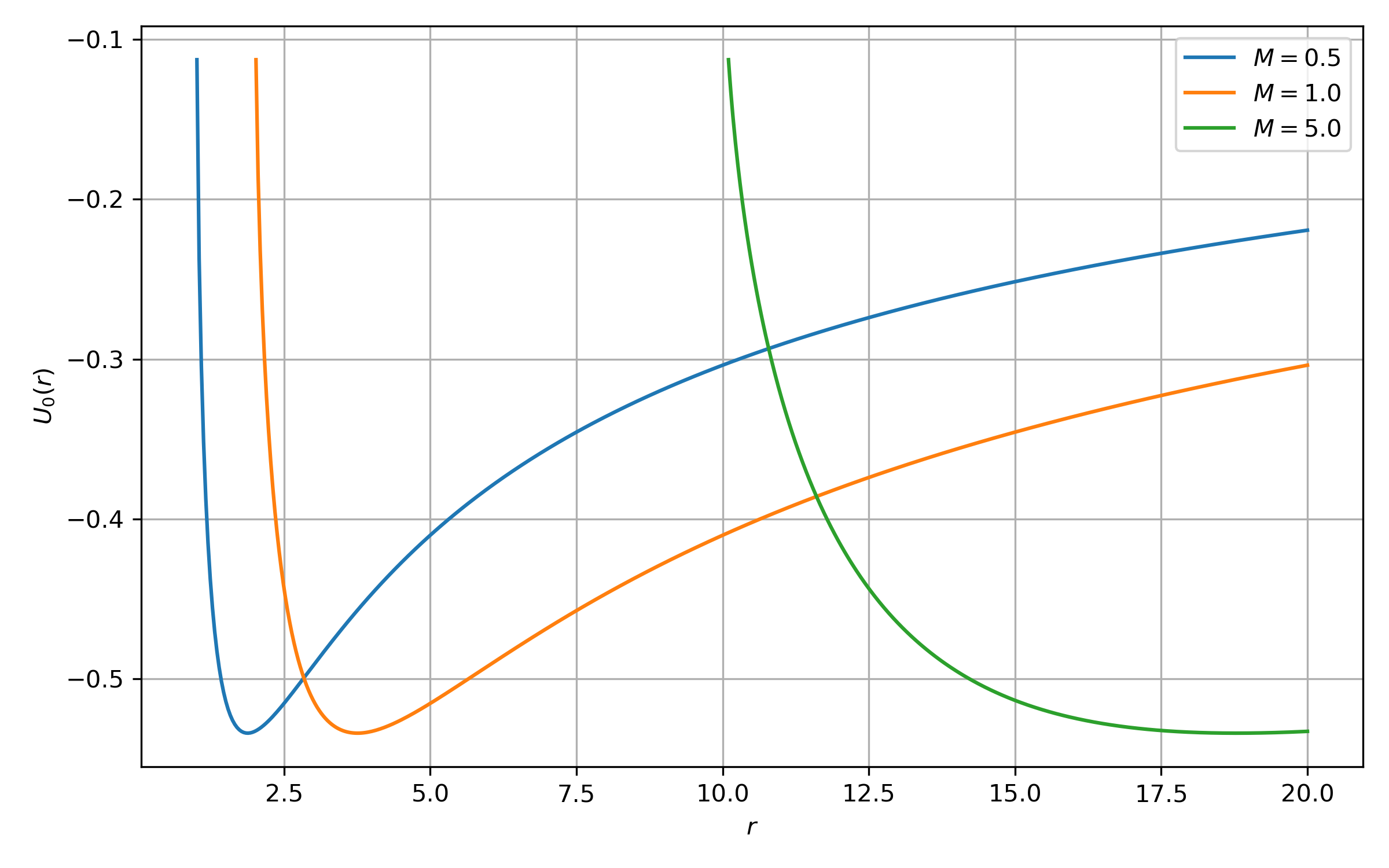}
    \caption{Inward stationary background flow velocity $U_{0}$ vs $r$ for different black hole mass $M$ }
    \label{fig:enter-theoq2}
\end{figure}

Furthermore, for simplicity, we assume a steady flow ${U}_0$ with $ \dot {U}_0 = 0 $ leading to a space-dependent background flow satisfying 
\begin{equation}
\partial_r(\frac{U_0^2}{f}) +\frac{B}{r^2}\frac{U_0^2}{f} +\frac{B}{r^2} =0.
    \label{l2}
\end{equation}
Exploiting the steady flow condition, $\dot\rho_0=0$, from the continuity equation we find
\begin{equation}
\partial_r (f \rho_0 U_0) = 0
\Rightarrow
f\rho_0 U_0 = \chi
\label{ll1}
\end{equation}
where $\chi$ is a constant. Thus, once the stationary flow $U_0(r)$ is derived from (\ref{l2}), the stationary density profile $\rho_0$ can be determined.
\\
{\it{The stationary background flow}}: In the $\dot{U}_0 = 0$ stationary background, from (\ref{l2}), with $f(r) = 1 - \frac{B}{r}, \quad B = 2GM$, defining 
\[
y(r) \equiv \frac{U_0^2}{f} \rightarrow \frac{dy}{dr}=\frac{d}{dr}\left(\frac{U_0^2}{f}\right) 
\]
we obtain
\[
\frac{dy}{dr} + \frac{B}{r^2} y = -\frac{B}{r^2}\rightarrow y(r) = 1 + C\,\exp\left(\frac{B}{r}\right)
\]
leading to
\[
U_0^2 = f\,\left[
1 + C\,\exp\left(\frac{B}{r}\right)
\right].
\]
Applying the boundary condition that $U_0\to0$ \cite{bondi} as $r\to\infty$, we recover the final result
\[
U_0(r) = \pm\sqrt{\left(1 - \frac{B}{r}\right)\left[-1 + \exp\left(\frac{B}{r}\right)\right]}.
\]
The positive and negative signatures refer to outward and inward background flows, respectively, with their respective velocity profiles depicted in Figure [\ref{fig:enter-theoq1}] and Figure [\ref{fig:enter-theoq2}].

\begin{figure}
    \centering
    \includegraphics[width=0.9\linewidth]{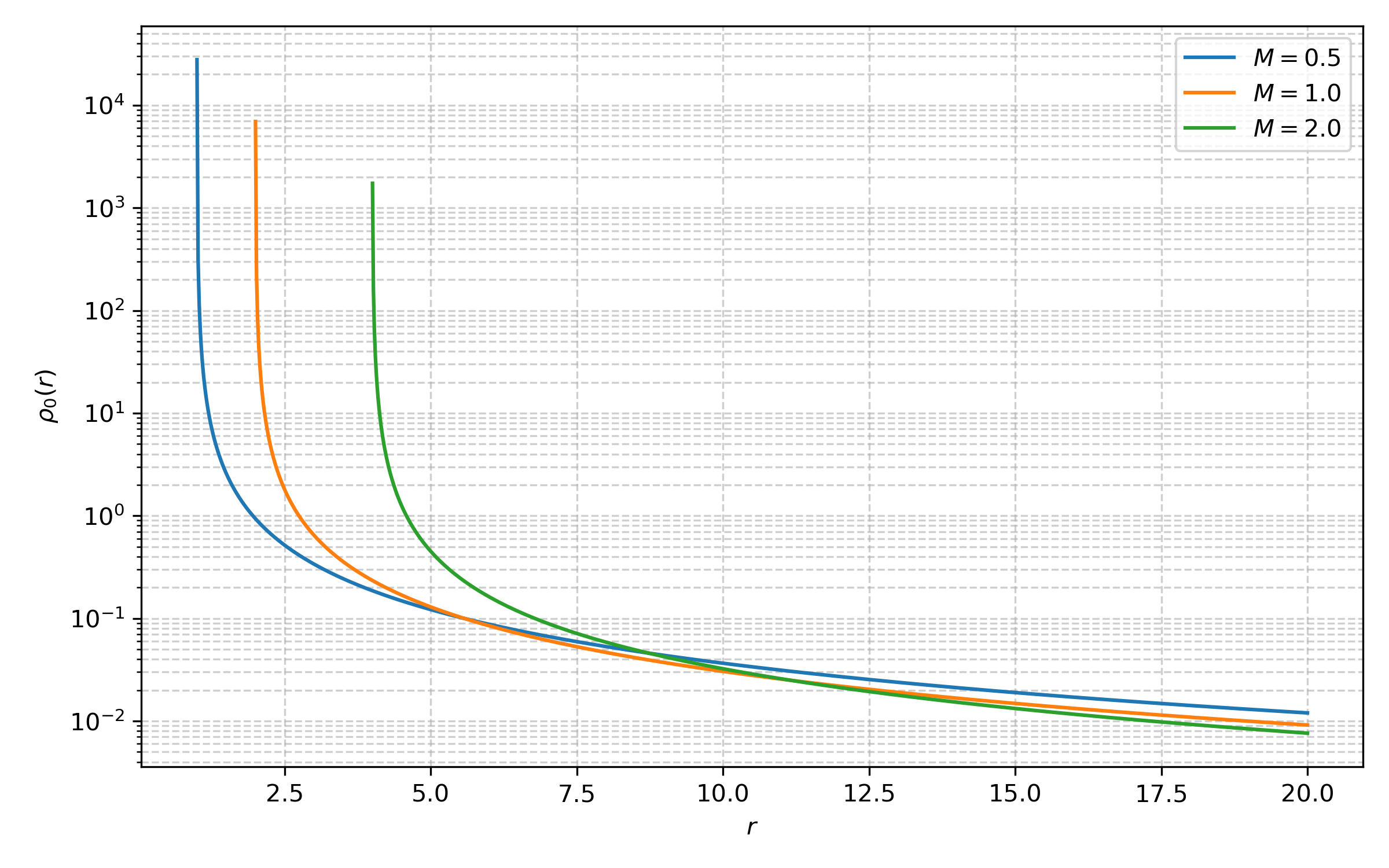}
    \caption{Inward/Outward stationary background flow density $\rho_{0}$ vs $r$ for }
    \label{fig:enter-theoq3}
\end{figure}

Coming to the stationary background density $\rho_0(r)$, as mentioned earlier in (\ref{ll1}), it is directly given by
\begin{equation}
\rho_0  = \frac{\chi}{fU_0}.
    \label{lll1}
\end{equation}
The corresponding (identical) profiles for inward and outward flows are given in Figure [\ref{fig:enter-theoq3}]. 

The background velocity and density are correlated through (\ref{lll1}), and this is reflected in the velocity and density graphs. In both outward and inward cases, the magnitude of the fluid velocity appears to be very small. In the former, the velocity drops smoothly to zero. On the other hand, the incoming flow velocity increases as the fluid nears the horizon and then decreases very close to the horizon. The {\it{coordinate}} velocity of fluid, vanishing at the horizon, is a manifestation of the freezing phenomenon;  matter takes infinite coordinate time to reach the horizon, which, however, is a finite proper time interval. As advertised throughout this paper, clearly this is the behaviour expected from matter close to the black hole horizon. This concludes our discussion on the stationary radial background fluid dynamics. In the next section, we will concentrate on the stability of this flow to first order in velocity and density perturbations.

\section{Fluid dynamics (in coordinate time) in Hamiltonian framework: \\
perturbations}
Let us now introduce perturbations \cite{An2024, VillaRampf2015,Friedman2013,DiazGuerra2024, DafermosReview, Benomio2022} (to be treated in first order) to the stationary velocity $U_0(r)$ and density $\rho_0(r)$ of the background flow discussed above.  
From (\ref{x11}, \ref{x12}, \ref{x13}) we know that the dynamical equations have a vorticity term, which, however, did not play any role in our background system that is restricted to be spherically symmetric in nature. As a consequence, only the background radial flow $U_0(r,t), \rho_0(r,t)$ was allowed and hence it was devoid of any vorticity. Hence, it will be interesting to study perturbations over this background flow to see how the angular fluid components are generated along with arbitrary coordinate dependence.  Thus, over the stationary radial background flow,
\begin{equation}
\rho_0(r),~~ U_0(r),~~ V=0,~~ W=0
    \label{p1}
\end{equation}
we introduce perturbations as defined below,
\begin{equation}
\rho =\rho_0(r) +\delta \rho (r,\theta,\phi,t),~~ U=U_0(r)+u(r,\theta,\phi,t),$$$$
V=v(r,\theta,\phi,t),~~W=w(r,\theta,\phi,t) .
    \label{p1}
\end{equation}
Note that the perturbations are allowed to depend on all the coordinates and time. In the present work, we will consider only first-order perturbations that are higher than linear order or products of perturbations will be dropped.
The density   perturbation equation to  first order is given by
\begin{equation}
(\delta\rho\dot  )+\frac{1}{r^2}\partial_r[r^2f(\rho_0u+U_0\delta \rho)] +\frac{1}{rsin~\theta}\partial_\theta (sin\theta~ f\rho_0v)+ \frac{1}{rsin~\theta}\partial_\phi (f\rho_0 w) =0 
    \label{p1}
\end{equation}
which, for the case of radial stationary background, turns into 
\begin{equation}
(\delta\rho\dot  )+\frac{1}{r^2}\partial_r[r^2f(\rho_0u+U_0\delta \rho)] +\frac{f\rho_0}{rsin~\theta}\partial_\theta (sin\theta~ v)+ \frac{f\rho_0}{rsin~\theta}\partial_\phi ( w) =0 .
    \label{pp1}
\end{equation}
Using $Z\approx f^{-1}(U_0^2+2U_0u)$ and keeping in mind that $U_0(r)$ is only $r$-dependent, the velocity perturbation equation to  first order are given by 
\begin{equation}
\dot u=-\,f\,U_{0}\,\partial_{r}u\, - u\,f\,\partial_{r}U_{0}
    \label{p2}
\end{equation}
\begin{equation}
\dot v=-\frac{U_0}{r^2}\partial_\theta u +\,U_0\,(\frac{1}{r^2}\partial_\theta u-f\partial_r v) 
    \label{p3}
\end{equation}
\begin{equation}
\dot w= -\frac{U_0}{r^2sin^2\theta}\partial_\phi u +\,U_0\,(\frac{1}{r^2sin^2\theta}\partial_\phi u-f\,\partial_r w) 
    \label{p4}
\end{equation}
In the above, we have used the fact that the vorticity $\Omega^{ij}=g^{ik}\partial_kU^j-g^{jk}\partial_kU^i$ with $U^r=U_0(r)+u,~U^\theta=v,~U^\phi=w$,
will comprise only perturbation terms, and following our convention, we switch to $\Omega^{ij}\rightarrow \omega^{ij}$ we obtain
\begin{equation}
\omega^{r\theta}=g^{rr}\partial_rv-g^{\theta\theta}\partial_\theta u,~\omega^{r\phi}=g^{rr}\partial_rw-g^{\phi\phi}\partial_\phi u,~\omega^{\theta\phi}=g^{\theta\theta}\partial_\theta w-g^{\phi\phi}\partial_\phi v.
    \label{p5}
\end{equation}
It is interesting to note three points: (i) the $u$-equations is decoupled and (ii) that $v,w$-equations are coupled and (iii) that the vorticity comes into play only in  $\dot v$ and $\dot w $ equations.

\begin{figure}
    \centering
    \includegraphics[width=0.9\linewidth]{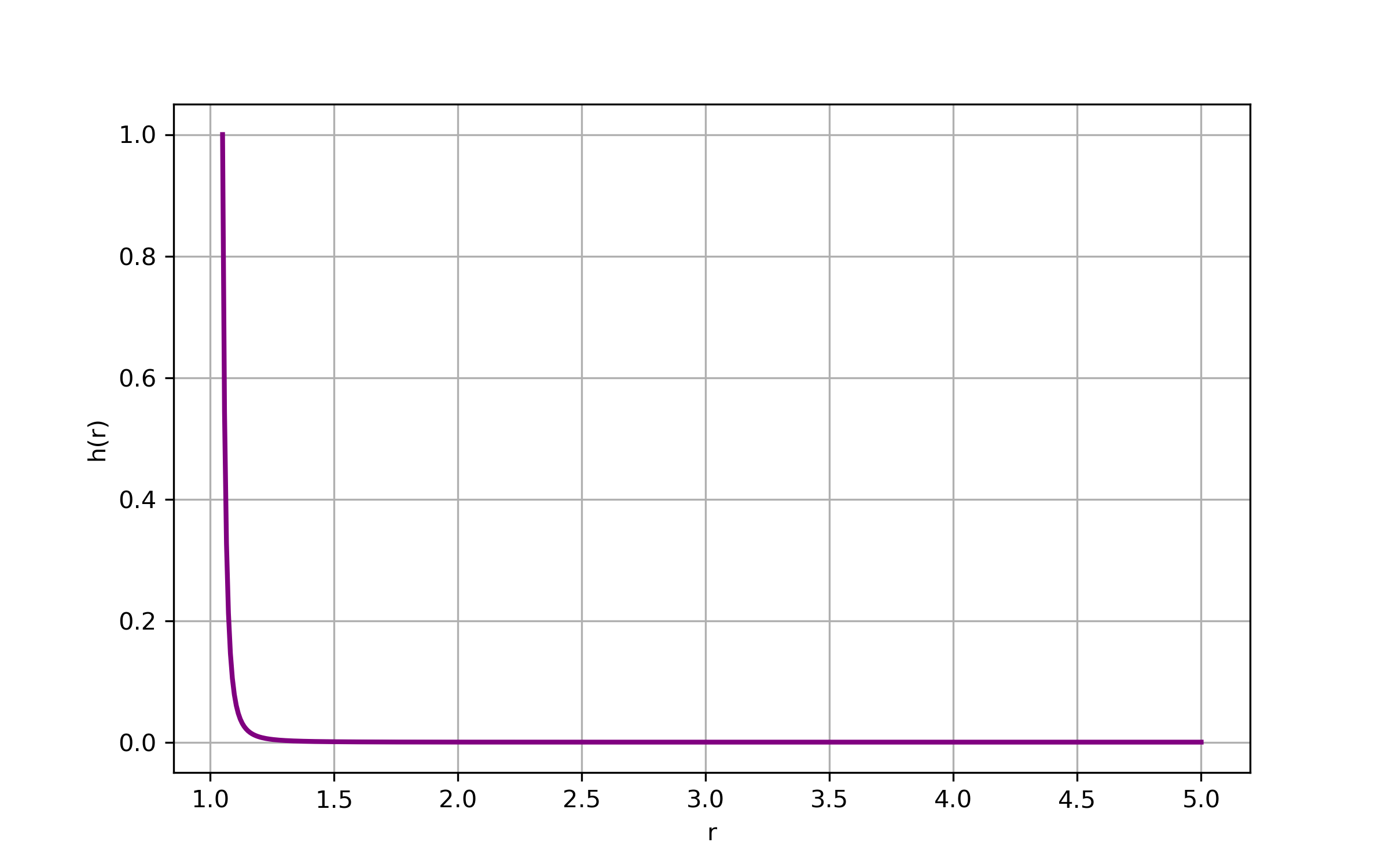}
\caption{Spatial part of radial perturbation  $u=h(r)\exp({-\kappa t})$ vs $r$}
    \label{fig:enter-theoq}
\end{figure}

\begin{figure}
    \centering
    \includegraphics[width=0.9\linewidth]{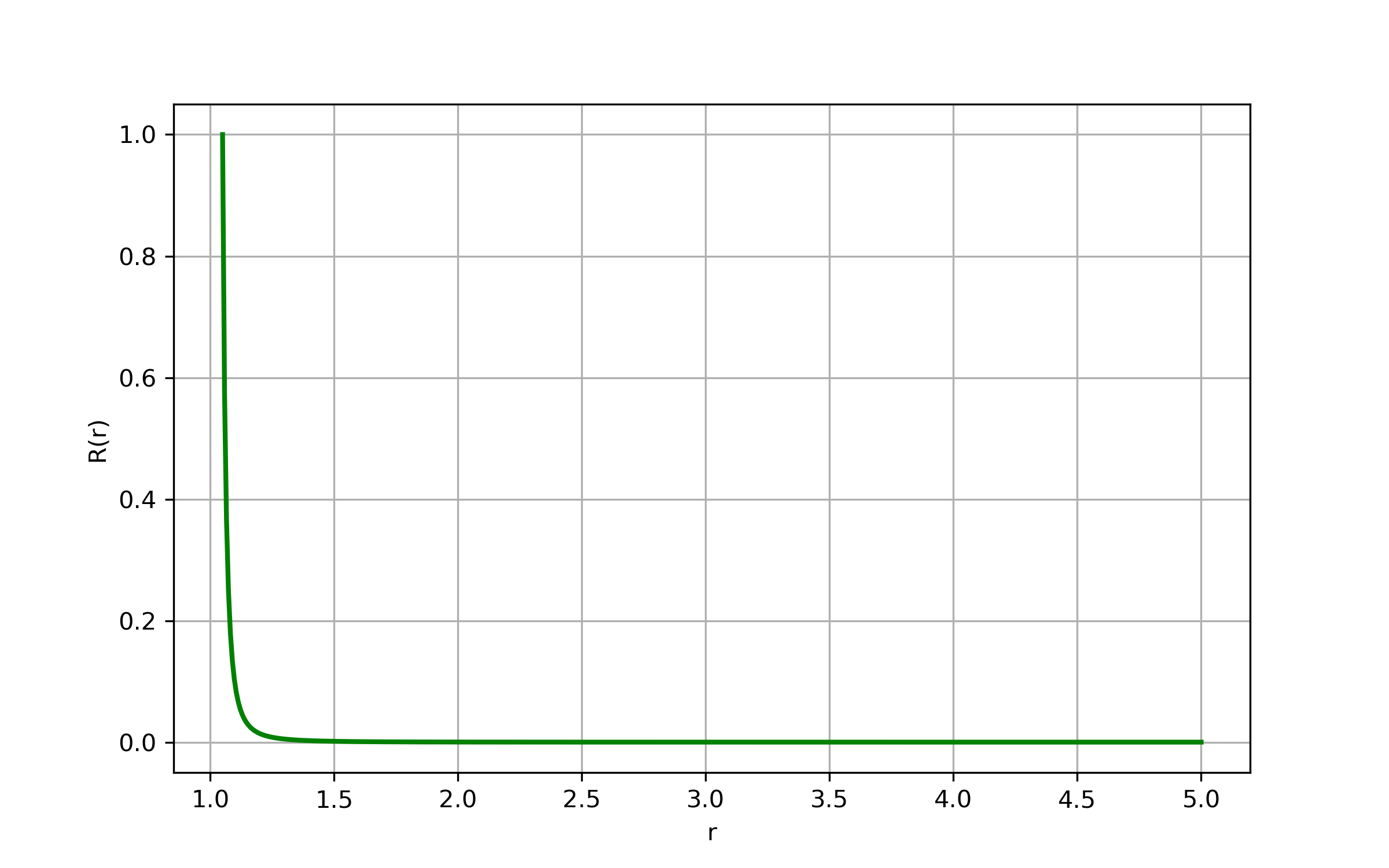}
    \caption{Spatial part of angular perturbation  $\omega(r)=R(r)exp(-\lambda t)$}
    \label{fig:enter-theoq8}
\end{figure}

\begin{figure}
    \centering
    \includegraphics[width=0.9\linewidth]{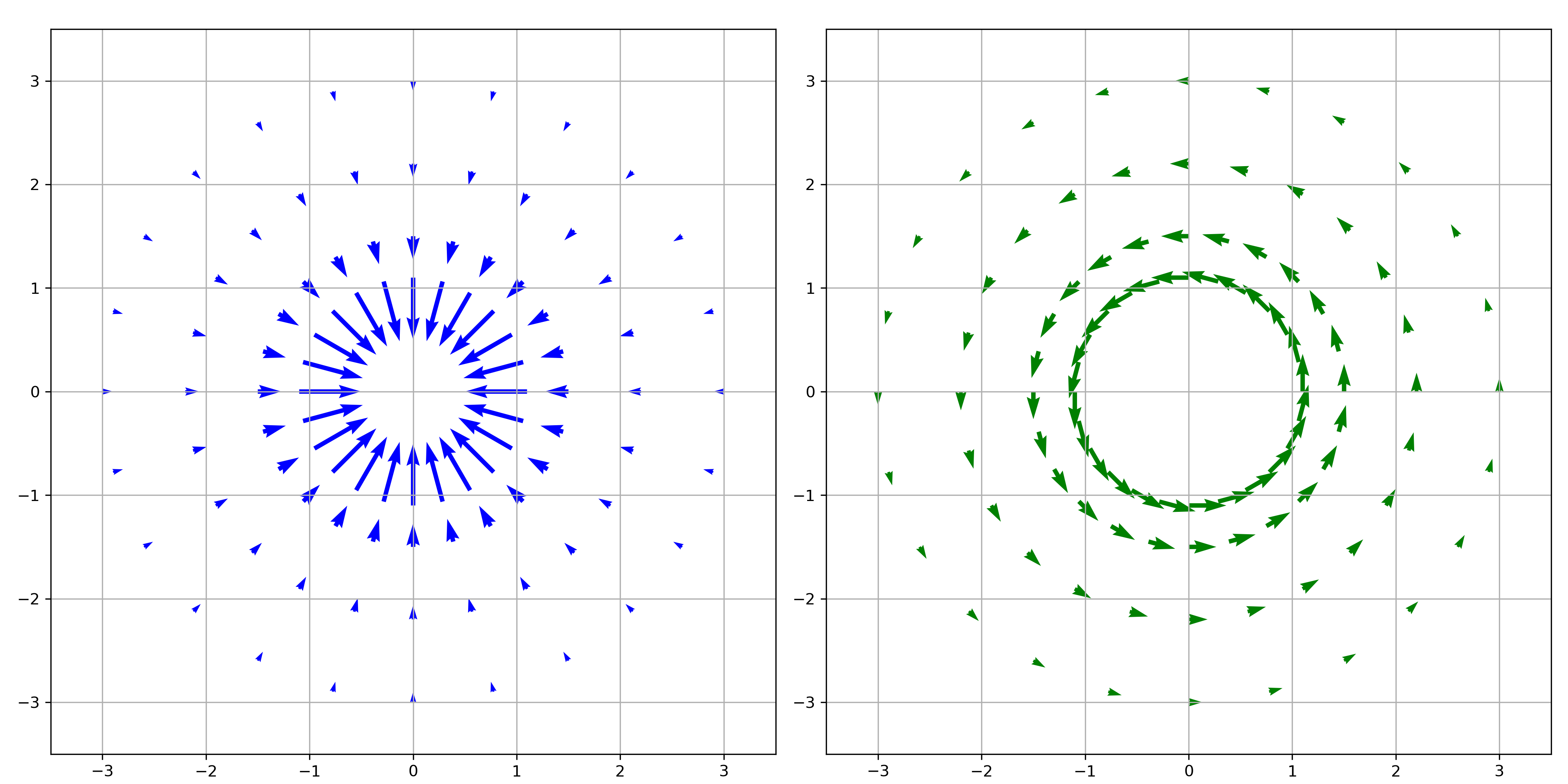}
    \caption{Spatial part of perturbation radial and angular  velocity}
    \label{fig:enter-theoq10}
\end{figure}

\begin{figure}
    \centering
    \includegraphics[width=0.9\linewidth]{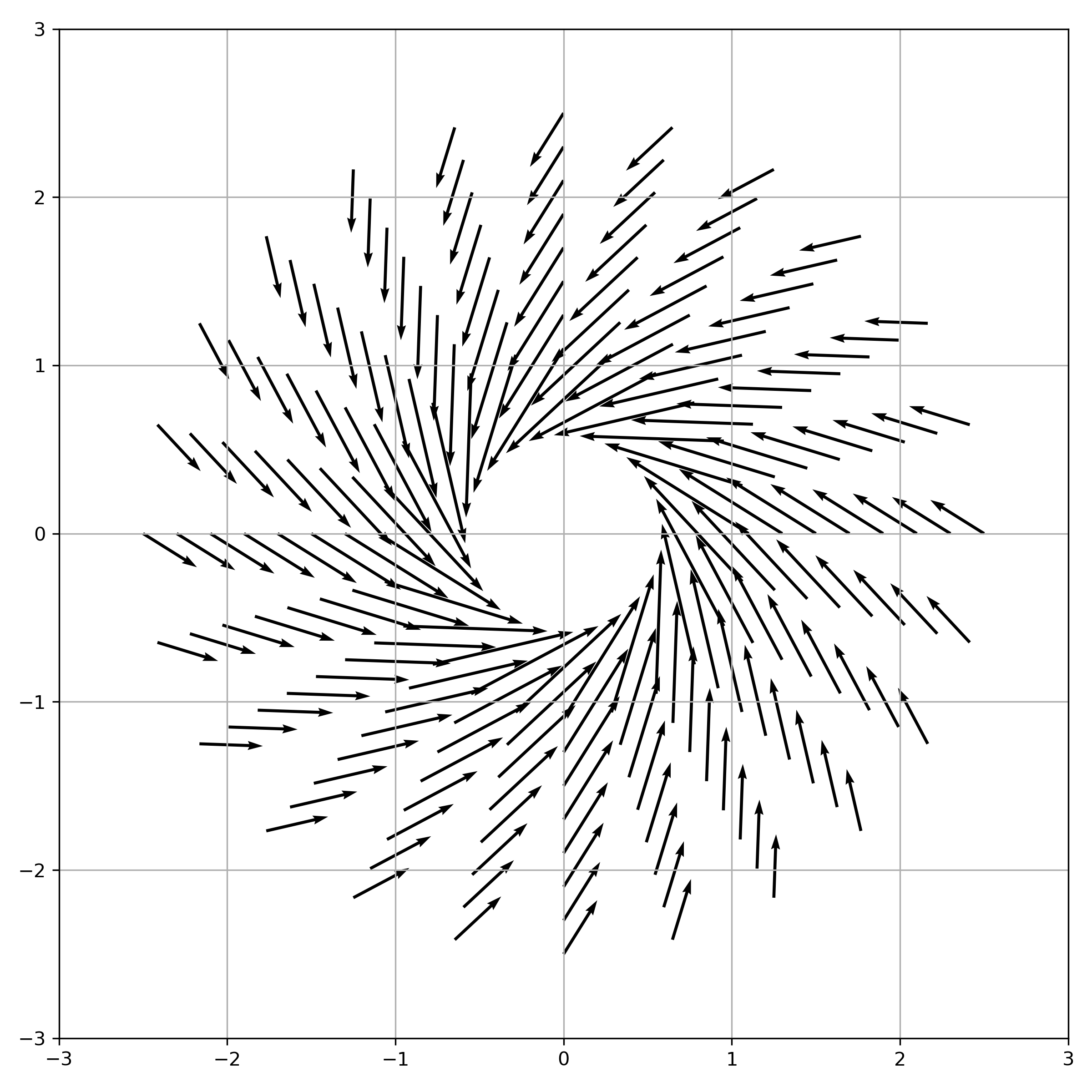}
    \caption{Spatial part of total velocity perturbation}
    \label{fig:enter-theoq71}
\end{figure}

{\it{Radial perturbation $u(r,t)$}}: Invoking separation of $t$ and $r$ variables, $u(r,t)$ is expressed as  $u(r,t)=h(r)\exp({-\kappa t})$  with $h(r)$ satisfying
\begin{equation}
\frac{U_0}{h}\,\frac{\partial h}{\partial r} + \frac{\partial U_0}{\partial r} = \frac{\kappa }{f}.
    \label{z1}
\end{equation}
Recalling an inward background  $U_{0}$  given by,
\[
U_0(r) =- \sqrt{\left(1 - \frac{B}{r}\right)\left(
-1 + \exp\left(\frac{B}{r}\right)
\right)}
\]
$h(r)$ is computed as  

\[
h(r)
=
C\,\exp\Biggl[
\int \left(
\frac{\kappa }{f(r)\,U_0(r)}
\;-\;
\frac{1}{U_0(r)}\,\frac{dU_0}{dr}
\right)
\,dr
\Biggr],
\]
or more explicitly
\begin{equation}
\label{h_sol}
h(r)
=
C\,\exp\Biggl\{
\int \frac{1}{U_0(r)}\Biggl[
\frac{k}{f(r)}
\;-\;
\frac{1}{2\,U_0(r)}\Bigl(
\frac{B}{r^2}\,\bigl[-1 + \exp\bigl(\tfrac{B}{r}\bigr)\bigr]
\;-\;
f(r)\,\frac{B}{r^2}\,\exp\bigl(\tfrac{B}{r}\bigr)
\Bigr)
\Biggr]
\,dr
\Biggr\}.
\end{equation}
The RHS can not be integrated to an analytic form and in Figure [\ref{fig:enter-theoq}]  we plot the profile numerically.\\
{\it{Angular perturbation $\omega(r,t)$}}: Without loss of generality, we can fix $\theta =\pi/2$, and focus on the $r,\phi$-plane where ${\bf{\omega}}$ satisfies \eqref{p4}. Again separating time and space dependence and writing $\omega(r,t)=g(r)exp(-\lambda t)$, we need to integrate
\begin{equation}
g(r) \;=\; C\,\exp\Biggl[
\lambda \,\int \frac{dr}{U_0(r)\,f(r)}
\Biggr].
\label{z2}
\end{equation}
We invoke numerical integration to generate the profile in Figure [\ref{fig:enter-theoq8}].
Let us draw cartoons to pictorially show the $r$-dependent part of radial $h(r)$ and angular $g(r)$ perturbations in the $ r-\phi$ $-$ plane, where the lengths of the arrows indicate the magnitude at that point. In Figure [\ref{fig:enter-theoq10}], we provide separate visuals of the radial and angular perturbations corresponding to an incoming stationary radial flow, and in Figure [\ref{fig:enter-theoq71}], the perturbation velocity components are combined to establish the vortex nature generated in the perturbation.

It is worthwhile to mention two important aspects of our analysis:\\
(I) The perturbations are exponentially decaying in time, indicating that the stationary, spherically symmetric background flow considered by us is stable.\\
(II) Recall that we had started with the perturbations having dependence on all the coordinates $r,\theta,\phi$. However, the specific nature of the background flow and first-order perturbation theory show that from (\ref{p2}), only the $r$-dependent part of the radial perturbation $u(r,\theta,\phi)$ can be determined. As the simplest choice, we have considered $u$ to depend only on the radial coordinate $r$. But once this choice on $u(r)$ is made, from (\ref{p4}) it is clear that on the $r-\phi$-plane, $\phi$-dependence of   $\omega$ (the angular part of the perturbation) remains undetermined.

\section{Conclusion and future directions}
In this paper, we have developed ideal fluid dynamical equations (continuity and Euler equations), applicable to a curved spacetime in a relativistic regime. We have worked in the Hamiltonian framework where the equations of motion are identified with   Poisson-like brackets between Eulerian density and velocity fields and a Hamiltonian function. This derived bracket structure and the posited Hamiltonian are our primary new and major results. It should be stressed that the dynamics unfolds in Eulerian (or coordinate) time from the perspective of an Eulerian observer, positioned at a fixed spacetime point. The continuity and Euler equations are derived, applicable to any arbitrary metric.

Subsequently, we specialise to Schwarzschild spacetime where the behaviour of an ideal fluid is studied in detail, especially near the event horizon. For simplicity, we choose a spherically symmetric background flow. The velocity and density profiles exhibit correct behaviour, thereby establishing the robustness of our generalised Hamiltonian framework.

Lastly, we have considered the general form of perturbations in density and velocity, which are not restricted by spherical symmetry. The perturbations, to first order, show two important features of our model: (I) the modes decay exponentially in time, indicating that the spherically symmetric background flow is stable, and (II) vorticity effects can be induced in first-order perturbations even though the background flow is free of vorticity.

Let us elaborate the above comments a little more. In this work, we have constructed a novel Hamiltonian formulation of relativistic fluid dynamics in arbitrarily curved spacetimes, working explicitly in Eulerian (coordinate) time. By retaining the familiar fluid variables—mass density and three-velocity—throughout, and rigorously extending the Poisson bracket structure and fluid Hamiltonian from flat spacetime to its curved spacetime counterpart, we have derived generalised continuity and Euler equations directly in a Hamiltonian framework. The formulation not only preserves the canonical structure of fluid dynamics but also remains physically transparent, especially from the viewpoint of static observers—a significant advantage over the conventional covariant formulations based on energy-momentum conservation.

Specialising to the Schwarzschild background, we obtained explicit expressions for a steady, spherically symmetric radial flow and investigated its perturbative stability. Our analysis shows that first-order perturbations—both radial and angular—decay exponentially in coordinate time, establishing the linear stability of the background solution. Furthermore, we demonstrate that vorticity, while absent in the background flow, emerges naturally at linear order in perturbations and plays a crucial role in the angular dynamics.

Our framework lays the groundwork for several promising extensions. These include applying the formalism to rotating (stationary but non-static) spacetimes such as Kerr, incorporating pressure and more general equations of state, studying nonlinear perturbations, and exploring possible connections to analogue gravity systems \cite{BarceloVisserReview}. The approach also offers potential advantages in numerical relativity and astrophysical modelling, particularly in problems involving accretion and horizon-scale fluid dynamics. The flexibility and physical transparency of the coordinate-time Hamiltonian framework developed here make it a compelling alternative for probing relativistic fluid behaviour in strong gravity regimes.

{\bf{Outlook and Future Work:}}
This work establishes a foundation for several promising extensions:\\
    Generalisations to Kerr spacetime: Incorporating frame-dragging effects for rotating black holes.\\
    Non-ideal fluids: Extending the formalism to include viscosity, heat conduction, and turbulence.\\
    Numerical implementation: Developing Hamiltonian-based simulations for relativistic accretion flows.\\
    Coupling to gravitational perturbations: Investigating backreaction effects in the fluid-spacetime system.\\
Our results demonstrate that the Hamiltonian approach offers a powerful and self-consistent framework for studying relativistic fluids in strong gravity, complementing traditional energy-momentum tensor methods. The explicit connection to non-relativistic fluid mechanics makes this formalism particularly suitable for bridging numerical relativity with established computational fluid dynamics techniques.\\
This work has direct relevance to black hole accretion, neutron star physics, and other high-energy astrophysical systems where fluid dynamics in strong-field gravity plays a crucial role. Future applications may include modeling horizon-scale dynamics in Event Horizon Telescope observations and studying instabilities in relativistic jets.

\newpage

\section {Appendices}
\begin{align}
   \dot{U}^r =\ & \frac{ \sqrt{-g_{00}} }{2 \Phi }\, g^{rr}  \bigg[
(\partial_r g_{00})(1 + Z) \nonumber \\
& + g_{00} \bigg{(} (\partial_r g_{rr})(U^r)^2 + (\partial_r g_{\theta\theta})(U^\theta)^2 + (\partial_r g_{\phi\phi})(U^\phi)^2 . \nonumber \\
& \quad +\, 2 g_{rr} U^r \partial_r U^r + 2 g_{\theta\theta} U^\theta \partial_r U^\theta + 2 g_{\phi\phi} U^\phi \partial_r U^\phi \bigg{)}
\bigg] \nonumber \\
& - \frac{g_{00}}{ \sqrt{1 + Z} } \left( \omega^{r\theta} g_{\theta\theta} U^\theta + \omega^{r\phi} g_{\phi\phi} U^\phi \right),
\end{align}

\begin{align}
\dot{U}^\theta =\ & \frac{ \sqrt{-g_{00}} }{2 \Phi }\, g^{\theta\theta}  \bigg[
(\partial_\theta g_{00})(1 + Z) \nonumber \\
& + g_{00} \bigg{(} (\partial_\theta g_{rr})(U^r)^2 + (\partial_\theta g_{\theta\theta})(U^\theta)^2 + (\partial_\theta g_{\phi\phi})(U^\phi)^2 
\nonumber \\
&. \quad +\, 2 g_{rr} U^r \partial_\theta U^r + 2 g_{\theta\theta} U^\theta \partial_\theta U^\theta + 2 g_{\phi\phi} U^\phi \partial_\theta U^\phi \bigg{)}
\bigg] \nonumber \\
& - \frac{g_{00}}{ \sqrt{1 + Z} } \left( \omega^{\theta r} g_{rr} U^r + \omega^{\theta\phi} g_{\phi\phi} U^\phi \right),
\end{align}

\begin{align}
\dot{U}^\phi =\ & \frac{ \sqrt{-g_{00}} }{2 \Phi }\, g^{\phi\phi}  \bigg[
(\partial_\phi g_{00})(1 + Z) \nonumber \\
& + g_{00} \bigg{(} (\partial_\phi g_{rr})(U^r)^2 + (\partial_\phi g_{\theta\theta})(U^\theta)^2 + (\partial_\phi g_{\phi\phi})(U^\phi)^2  \nonumber \\
&  \quad +\, 2 g_{rr} U^r \partial_\phi U^r + 2 g_{\theta\theta} U^\theta \partial_\phi U^\theta + 2 g_{\phi\phi} U^\phi \partial_\phi U^\phi \bigg{)}
\bigg] \nonumber \\
& - \frac{g_{00}}{ \sqrt{1 + Z} } \left( \omega^{\phi r} g_{rr} U^r + \omega^{\phi\theta} g_{\theta\theta} U^\theta \right).
\end{align}

\[
\dot{\rho}
\;+\;
\frac{1}{r^2}\,\frac{\partial}{\partial r}\Biggl[
r^2\,
\frac{\,f(r)\,\rho\,}{\sqrt{
\,1 + \dfrac{(U^r)^2}{f(r)} + r^2 (U^\theta)^2 + r^2 \sin^2\theta\, (U^\phi)^2
}}
\,U^r
\Biggr]
\]

\[
+
\frac{1}{r\,\sin\theta}\,\frac{\partial}{\partial\theta}\Biggl[
\sin\theta\,
\frac{\,f(r)\,\rho\,}{\sqrt{
\,1 + \dfrac{(U^r)^2}{f(r)} + r^2 (U^\theta)^2 + r^2 \sin^2\theta\, (U^\phi)^2
}}
\,U^\theta
\Biggr]
\]

\begin{equation}
+
\frac{1}{r\,\sin\theta}\,\frac{\partial}{\partial\phi}\Biggl[
\frac{\,f(r)\,\rho\,}{\sqrt{
\,1 + \dfrac{(U^r)^2}{f(r)} + r^2 (U^\theta)^2 + r^2 \sin^2\theta\, (U^\phi)^2
}}
\,U^\phi
\Biggr]
=0.
\end{equation}

\end{document}